\documentclass[apj]{emulateapj}
\usepackage{epsfig,rotating,amsmath,natbib}
\bibdata{msnew}
\shorttitle{HI in Local Group Dwarfs}

\accepted{to ApJ on 1/12/09}
\begin{document}

\title{HI in Local Group Dwarf Galaxies and Stripping by the Galactic Halo}
\author{Jana Grcevich\altaffilmark{1},
        Mary E Putman\altaffilmark{1,2}}
\altaffiltext{1}{Department of Astronomy, University of Michigan, Ann Arbor, MI 48019}
\email{janagrc@gmail.com}
\altaffiltext{2}{Department of Astronomy, Columbia University, New York, NY 10027}
\email{mputman@astro.columbia.edu}

\def\spose#1{\hbox to 0pt{#1\hss}}
\def\simlt{\mathrel{\spose{\lower 3pt\hbox{$\mathchar"218$}}
     \raise 2.0pt\hbox{$\mathchar"13C$}}}
\def\simgt{\mathrel{\spose{\lower 3pt\hbox{$\mathchar"218$}}
     \raise 2.0pt\hbox{$\mathchar"13E$}}}
\def\etal       {{\it et al.}}
\def\gtrapprox  {\;\lower 0.5ex\hbox{$\buildrel >\over \sim\ $}}
\def\lessapprox {\;\lower 0.5ex\hbox{$\buildrel < \over \sim\ $}}
\def\Msun       {${\rm M}_\odot$}
\def\deg        {$^\circ$}
\def\sigLL      {\sigma_{\scriptscriptstyle Lyc}}
\def\nH         {n_{\scriptscriptstyle H}}
\def\NH         {$\rm N_{\scriptscriptstyle H}$}
\def\HI         {\ion{H}{1}}
\def\Ha         {${\rm H}\alpha$}
\def\eg         {{\it e.g.,\ }}
\def\ie         {{\it i.e.,\ }}
\def\cf         {{\it cf.\ }}
\def\qv         {{\it q.v.,\ }}
\def\kms        {km~s$^{-1}$}
\def\cmmsq      {cm$^{-2}$}

\def\plotfiddle#1#2#3#4#5#6#7{\centering \leavevmode
\vbox to#2{\rule{0pt}{#2}}
\includegraphics{#1}}

\begin{abstract}
We examine the HI content and environment of all of the Local Group dwarf galaxies (M$_{tot} < 10^{10}$ \Msun), including the numerous newly discovered satellites of the Milky Way and M31. 
All of the new dwarfs, with the exception of Leo T, have no detected HI.  The majority of dwarf galaxies within $\sim$270 kpc of the Milky Way or Andromeda are undetected in HI (\lessapprox $10^5$ \Msun~ for Milky Way dwarfs), while those further than $\sim$270 kpc are predominantly detected with masses  $\sim$10$^{5}$ to $10^{8}$ M$_{\odot} $. 
Analytical ram-pressure arguments combined with velocities obtained via proper motion studies allow for an estimate of the halo density of the Milky Way at several distances. 
This halo density is constrained to be greater than 2-3 $\times$ 10$^{-4}$ cm$^{-3}$ out to distances of at least 70 kpc.  This is broadly consistent with theoretical models of the diffuse gas in a Milky Way-like halo and is consistent with this component hosting a large fraction of a galaxy's baryons.  Accounting for completeness in the dwarf galaxy count, gas-less dwarf galaxies could have provided at most $2.1 \times 10^{8}$ M$_{\odot}$ of HI gas to the Milky Way, which suggests that most of our Galaxy's star formation fuel does not come from accreted small satellites in the current era.
\end{abstract}

\section{Introduction}

The Sloan Digital Sky Survey data \citep{york00} have led to the discovery of 13 new satellites of the Milky Way
\citep{belokurov06a, zucker06a, willman05b, willman05a, belokurov06b, irwin07,walsh07, zucker06b,belokurov08}. Similar objects have been found
in the vicinity of M31 \citep{zucker04a, martin06, zucker07, majewski07}. These satellites have absolute
magnitudes between those of globular clusters and dwarf spheroidal galaxies, with most resembling faint dwarf
spheroidals. The discovery of these satellites has implications for the nature of the smallest galaxies in the universe and the building up of larger galaxies.   In particular, the newly discovered satellites partially alleviate the ``missing satellites problem'', or the order of magnitude discrepancy between the dark matter haloes predicted by $\Lambda$CDM simulations and the number of dwarf galaxies observed in the Local Group \citep{klypin99, moore99, koposov07b}.  Knowledge of the composition of these satellites is important to determine the mechanisms responsible for the formation and evolution of the smallest galactic building blocks and the fuel they bring to larger galaxies like the Milky Way. In addition, the satellites can be used to probe the extended gaseous halo of the parent galaxy, a galactic component which is difficult to detect directly and may harbor a significant fraction of the galaxy's baryons \citep{sommerlarsen06, maller04}.

This paper presents a study of the gaseous environments of the dwarf galaxies of the Local Group. These dwarfs are found at a wide range
of galactocentric distances with most of the newly discovered dwarfs well within the Milky Way's dark matter halo ($\sim$200 kpc). The close proximity of the Milky Way satellites and their range in 
Galactocentric radii make them an excellent choice to study stripping and gas loss mechanisms. 
In this paper we use existing HI observations to put constraints on the neutral gas content of Local Group dwarf galaxies and use this information to
examine what dictates their gas content and probe the diffuse Galactic halo. This paper includes all newly discovered dwarfs through November 2008 as well as the previously known dwarf galaxies listed in \cite{mateo98} which have total masses less than 10$^{10}$ M$_{\odot}$.\footnote{The SMC and LMC are not included due to their large combined total mass and the complex 3-body interaction of this system.}

\section{Observations}

The HI data are from the HI Parkes All-Sky Survey (HIPASS: Barnes et al.~2001\nocite{barnes01}) and the Leiden/Argentine/Bonn Survey (LAB: Kalberla et al. 2005\nocite{kalberla05}). The HIPASS data cover the sky at $\delta <$ +25$^{\circ}$, and have
been reprocessed using the MINMED5 algorithm which increases the sensitivity to resolved structures by using the entire 8\deg~HIPASS scan to calculate the bandpass correction (rather than only using a 2\deg~section of the scan; see Putman et al. 2003 for details\nocite{putman03}). The reprocessed data range from -700 to 1000 km s$^{-1}$ with reference to the LSR and the main beam FWHM is 15.5$^{\prime}$ after
gridding the data into cubes.  The
channel spacing is 13.2 km s$^{-1}$, and the spectral resolution after Hanning smoothing is 26.4 km s$^{-1}$. The average
single channel rms noise level in the northern data ($\delta$ = 2$^{\circ}$ - 25$^{\circ}$) with known dwarf galaxies is 13 mK, and this corresponds to a 5$\sigma$ HI mass detection limit (using the 26.4 km s$^{-1}$ velocity resolution) of $M_{HI} = 0.5 \: D_{kpc}^{2} \: M_{\odot}$.  Though the southern HIPASS data have a lower noise level on average, we also use this limit for the limited sources that are in the southern data given the variation in noise level in HIPASS cubes \citep{zwaan04}.

The LAB survey, the Leiden/Argentine/Bonn Galactic HI Survey \citep{kalberla05}, covers the entire sky by combining the
Leiden/Dwingeloo survey (LDS: Hartmann \& Burton 1997\nocite{hartmann97}) and the Instituto Argentino de
Radioastronom\'{i}a Survey (IAR: Arnal et al. 2000; Bajaja et al. 2005\nocite{arnal00, bajaja05}). The data have also
been recorrected for stray radiation. The LAB survey spans velocities from -450 to 400 km s$^{-1}$ with a channel spacing of
1.3 km s$^{-1}$. The main beam FWHM is 35.7$^{\prime}$, and the rms brightness temperature noise ranges from 70 to 90 mK for a
single 1.3 km s$^{-1}$ channel.  Assuming a dwarf has a minimum velocity range of 10 km s$^{-1}$, and smoothing LAB to this width, the 5$\sigma$
detection limit as a function of distance is $M_{HI} = 2.5 \: D_{kpc}^{2} \: M_{\odot}$.   This assumes the source is relatively close to the center of the LAB pointing.

The HIPASS data represent an important improvement in sensitivity and spatial resolution as compared to the LAB data. While we are not the first to use HIPASS data to study the HI in dwarf galaxies (see Bouchard et al. 2006; hereafter BCS06\nocite{BCS06}), the environment of the newly discovered dwarfs have not been previously explored, and we are also able to set deeper limits on several other Local Group dwarf galaxies and collate the results of the entire sample. It is important to note that the HI mass limits obtained with the above are only valid if the size of the dwarf is smaller than the beam width, which is $\sim$36$^{\prime}$ for the LAB data and 15.5$^{\prime}$ for the HIPASS data. This corresponds to a physical size of 305 pc for the nearest dwarf with limits obtained from LAB data (Ursa Major II at 30 kpc) and 2.2 kpc for the furthest dwarf with a LAB limit (Canis Venetici I at 220 kpc).  Minimum physical sizes for confident nondetection of the galaxies in the HIPASS region range from 103 pc for Segue I at a distance of 23 kpc to 721 pc for Leo IV at a distance of 160 kpc. In all cases the half light radius of the dwarf is smaller than the beam width.

\section{Results}

\subsection{Non-detections} 
\subsubsection{Newly Discovered Milky Way Satellites}

We have examined the HIPASS and LAB data in the vicinity of the newly discovered satellites of the Local Group. In all the optical centers of the new satellites examined, except that of Leo T, there was no detection of an
HI cloud at the velocities of the dwarf galaxies. There was no evidence of HI near Leo V, Segue I, Coma Berenices, Bo\"{o}tes II, or Hercules in the HIPASS data, and we confirm the non-detection and improve the HI mass limit with HIPASS for Bo\"{o}tes I \citep{bailin07}. LAB data show no evidence of HI in Canis Venetici I or
II, Ursa Major I or II, or Willman I.  Upper limits for the HI masses of the undetected satellites are determined by the relevant 5$\sigma$ detection limit and the distance to the satellite; they range from 265 M$_{\odot}$ to 12 $\times$ 10$^{4}$ M$_{\odot}$ (see Table 1).  

The search for emission was completed within 100 \kms~of the optical velocity of the dwarf.   In some channels Galactic emission is present which interferes with the ability to detect emission from the dwarf.  This ``Galactic Interference Range'' is noted in Table 1.  For three of the dwarf galaxies with non-detections the optical velocity lies within the Galactic Interference Range. In these cases the diffuse Galactic emission causes a higher noise level, and in order to set accurate mass limits the noise is calculated at the optical velocity. In addition, Leo IV has an HVC in the vicinity which causes the limit to be higher.   The discrete nature of dwarf galaxy HI emission does generally make a detection rise out of the Galactic emission, as in the case of Leo T \citep{irwin07}. None of the new dwarfs show this type of discrete emission in their vicinity.

\subsubsection{New and Previously Known M31 Satellites}

The HI environment of both the previously known and newly discovered M31 satellites were examined.  This section describes only those satellites with non-detections that have been clearly defined as M31 satellites.  Upper limits for the HI mass were determined with LAB data for Andromeda III, IX, X, XI, XII, XIII, XIV, XV, XVI, M32 and XVII. We also confirm HI nondetections for Andromeda I, II, and VII with LAB data and And VI with HIPASS data (Blitz \& Robishaw 2000; hereafter BR00\nocite{BR00}). The velocity measured by \cite{harbeck01} of Andromeda V indicates that the detection in BR00 is false.

The HI upper limits for undetected M31 satellites are listed in Table 2. We list the $5\sigma$ detection limits for the LAB data, except for And VI, which is the $5\sigma$ HIPASS limit, and for NGC 147 the limit comes from other sources. Andromeda IV is not included because it is not associated with M31 \citep{ferguson00}, and Andromeda VIII is excluded because its existence is in dispute \citep{merrett06}. Andromeda XI, XII, and XIII have uncertain distances; for the purpose of calculating an HI mass upper limit they are assumed to be at the distance of M31 (784 kpc, Stanek \& Garnavich 1998\nocite{stanek98}), but the limits listed in Table 2 for these dwarfs are approximate and they are excluded from Figure~\ref{hidist}.

\subsubsection{Other Previously Known Local Group Dwarfs}

Prior studies have examined the HI environment of the previously known dwarf galaxies. In the case of Cetus, an HI cloud within 1.5$^{\circ}$
of the optical center was found at a velocity of -280 km s$^{-1}$ (BCS06). Since that time
the optical velocity of Cetus has been measured by \cite{lewis07} as being -87 km s$^{-1}$, so the cloud is not
associated with the galaxy. A detection of low significance near the position and optical velocity of the Sextans dwarf
galaxy was found using data with less sensitivity and resolution than the HIPASS data (BR00\nocite{BR00}), but no cloud was found after inspection of the HIPASS cubes. A cloud reported near the position of Leo I at a velocity 26 km s$^{-1}$ from the
optical velocity of the dwarf (BR00) was also not found upon inspection of the HIPASS data, in agreement with BCS06.

Three clouds were reported near the Carina galaxy by BCS06. Of the three clouds, two are near the optical edge of
the galaxy at a distance of about 80$^{\prime}$ (2.3 kpc) from the optical center, and have velocities close to the optical velocity of Carina. There is no HI within the optical radius of the galaxy and the clouds lie outside the tidal radius, so
it is unlikley that the gas and the dwarf are physically associated (BCS06).

HI in the general direction of Tucana was first detected by \cite{oosterloo96} who claimed it was associated with the
Magellanic Stream. BCS06 also detect this cloud at a velocity of about 130 km s$^{-1}$, and offset from central position of the dwarf by $\sim$ 18$^{\prime}$. The optical velocity for Tucana has been found by \cite{tolstoy04} to be 182 km s$^{-1}$, so the difference in velocity between the HI cloud and the optical
dwarf is 52 km s$^{-1}$. Figure~\ref{ambig} shows the average velocity along the line of sight with the integrated intensity contours overlaid for the vicinity of the Tucana Dwarf. The optical position of the dwarf is marked with a plus sign. Due to its proximity to the Magellanic Stream and the offset velocity of the cloud near the Tucana dwarf, we consider it a non-detection.

To summarize our findings for the non-detections of previously known Local Group dwarfs, we confirm the non-detections or improve HI mass limits with HIPASS for Leo I and Leo II \citep{knapp78}, Cetus, Carina, Sextans (BCS06\nocite{BCS06}), and the core of the Sagittarius stream \citep{koribalski94} and find that the gas in the vicinity of Tucana is unlikely to be associated. We confirm non-detections in the LAB data for Ursa Minor and Draco \citep{knapp78}, and NGC 147 \citep{young97}. The HI limits for previously known Local Group dwarfs except the Andromeda dwarfs are given in Table 3.

\subsection{Ambiguous Detections} 
\begin{figure*}
 \epsscale{75.0} 
\includegraphics[width=7 in]{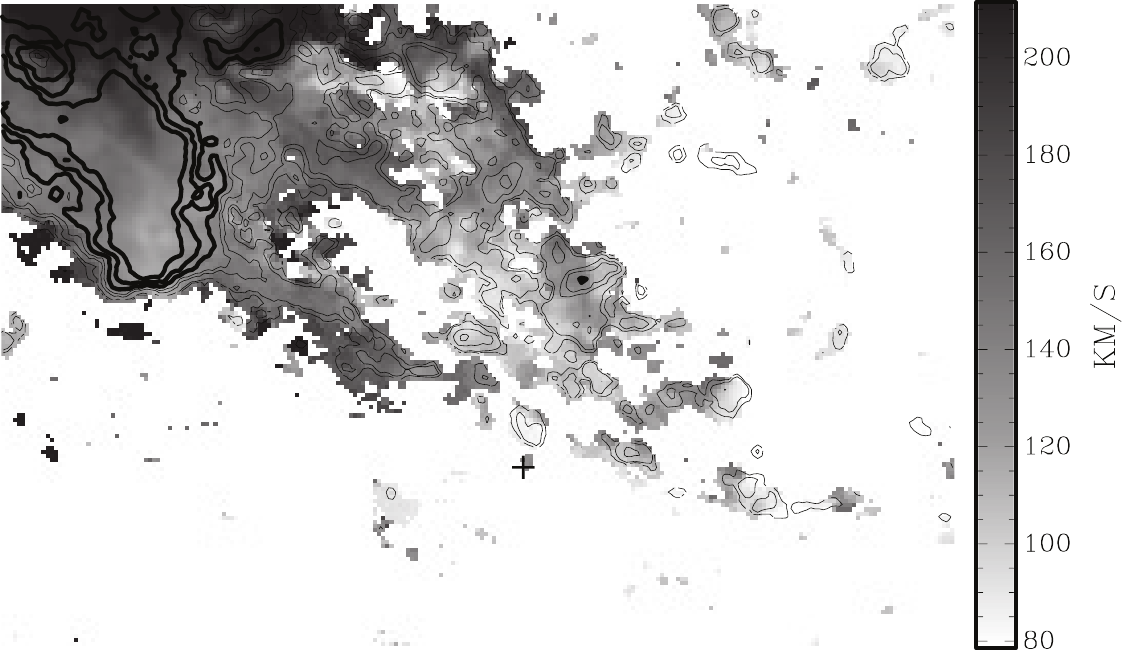}\\
\\
\\
\\
 \includegraphics[width=3 in]{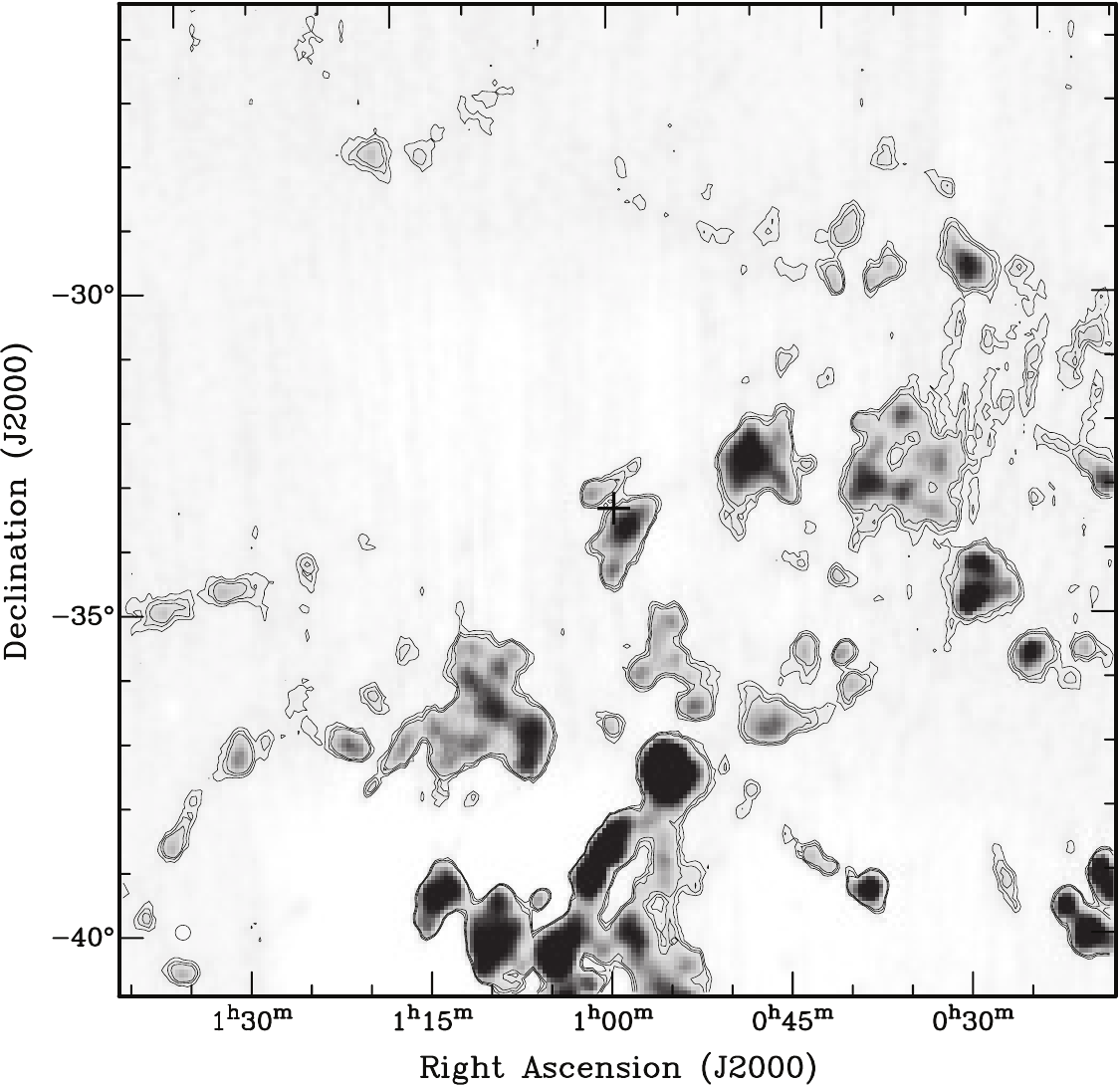}
 \hspace{20 mm}
 \includegraphics[width=3 in]{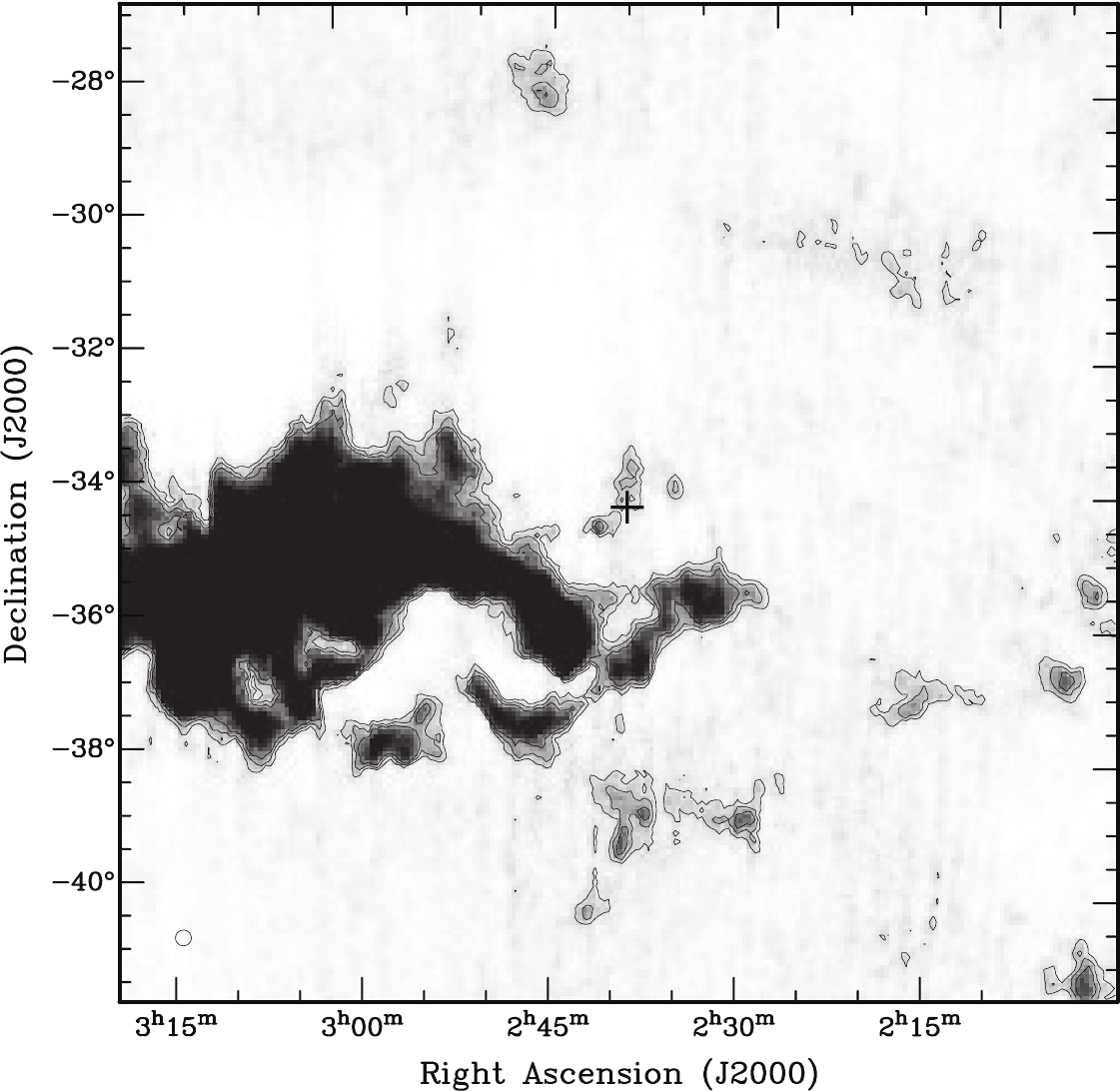}
 \caption{The average velocity along the line of sight with the integrated intensity contours overlaid for the region of Tucana (top), and integrated intensity maps for the regions of the Sculptor (bottom left) and Fornax (bottom right) dwarfs.  The optical positions of the dwarfs are marked with a plus sign.  The contours for the Tucana region are at 0.911, 1.82, 5.47, 9.11, 18.2, 36.5, and 72.9 $\times 10^{19}$ cm$^{-2}$. The contours are 0.50, 0.84, and 1.2 $\times 10^{19}$ cm$^{-2}$ for Fornax and 0.81, 1.3, and 1.9 $\times 10^{19}$ cm$^{-2}$ for Sculptor. The velocities included in the integrated intensity maps are 20 to 46 km s$^{-1}$ for Fornax and 46 to 152 km s$^{-1}$ for Sculptor, while the optical velocities are 53 and 102 km s$^{-1}$, respectively. }
 \label{ambig}
 \end{figure*}

In the case of two of the dwarfs, it is unclear if a cloud is associated with the
dwarf galaxy or has a separate origin. We refer to these as ambiguous detections, but note the mass of the cloud at the distance
of the galaxy in Table 3. One of these is the Sculptor Dwarf, near which two HI clouds were discovered by
\cite{carignan98}. The velocity of the HI ($\sim$ 105 km s$^{-1}$) and the optical velocity of the dwarf galaxy (102 km
s$^{-1}$) agree. Despite this, the Sculptor dwarf is in the same direction as the Magellanic Stream and another complex of HI clouds in the general direction of the Sculptor Group \citep{putman03}, and numerous clouds near that velocity are found in this region that could be mistaken for gas associated with the Sculptor dwarf (see Figure~\ref{ambig}). For these reasons, as well as the offset of the clouds from the optical center and the lack of recent star formation in this dwarf, we consider the Sculptor detection ambiguous.

Another ambiguous case is that of the Fornax dwarf. The Fornax dwarf has an optical velocity of 53 km s$^{-1}$. The cloud in question suffers from contamination from Galactic HI emission in the HIPASS data and is offset 28$^{\prime}$ from the optical center of Fornax. In the analysis by BCS06 the removal
of the Milky Way's spectrum was incomplete, and it was unclear if the cloud was part of the Milky Way, a cloud of
separate origin, or associated with the Fornax dwarf. We confirm that the origin of the cloud is unclear from the HIPASS data, which is shown in its environment in Figure~\ref{ambig}.

\subsection{Confident Detections} Clear detections of HI at the position and velocity of the dwarf have been made for Antlia, Phoenix, Pegasus, DDO 210, 
WLM, IC 5152, UGCA 438, LGS3, Sextans B, IC1613, Sextans A, GR8, Sagittarius, and SagDIG. We confirm all of these detections in the HIPASS data, although the Phoenix dwarf blends into Galactic emission. The HI masses of these and additional Local Group dwarfs as well as references are listed in Table 3.

The HI detection of LGS3
(Hulsbosch \& Wakker 1998; BCS06\nocite{hulsbosch88, BCS06}) is unusual in that it has one cloud at the optical
position of the dwarf and two clouds offset from the optical center which have diffuse HI connecting them. Only the cloud aligned with the position of LGS3 has a velocity which agrees closely with the optical velocity of LGS3, so only the mass of that cloud is considered.

Leo T is one of the newly discovered SDSS dwarf galaxies and is a particularly interesting object due to its low luminosity, recent history of star formation, and gas content. The HIPASS data show a compact HI cloud in the direction of the Leo T dwarf which was first reported
by \cite{irwin07}, and confirmed by \cite{ryanweber08} with synthesis data. Leo T has a velocity of about 35 km s$^{-1}$, and in the HIPASS data a maximum velocity of about 46 km s$^{-1}$. The
lower velocity cutoff and the total width in velocity are uncertain due to Galactic interference.
Our reanalysis of the HI cloud as it appears in the HIPASS data
reduced to recover extended emission indicates Leo T has a total HI mass of about 4.3 $\times$ 10$^{5}$
M$_{\odot}$ assuming a distance of 420 kpc.\footnote{The HI mass of Leo T was also checked with new data from the Galactic Arecibo L-band Feed Array (GALFA) HI Survey, which has a velocity resolution of 0.74 \kms  (e.g., Stanimirovic et al. 2006\nocite{stanimirovic06}), and found to be 4.8 $\times$ 10$^{5}$ M$_{\odot}$.}  \cite{ryanweber08} found a total HI mass of 2.8 $\times$ 10$^{5}$ M$_{\odot}$ and a peak HI column density of 7 $\times$ 10$^{20}$ cm$^{-2}$. Our higher total mass may be due to extended emission of Leo T not recovered in the synthesis maps and/or some level of diffuse Galactic emission included in the integrated intensity map.

\subsection{HI Mass and Galactocentric Radius} 
\begin{figure*}
\includegraphics[width=0.98\textwidth]{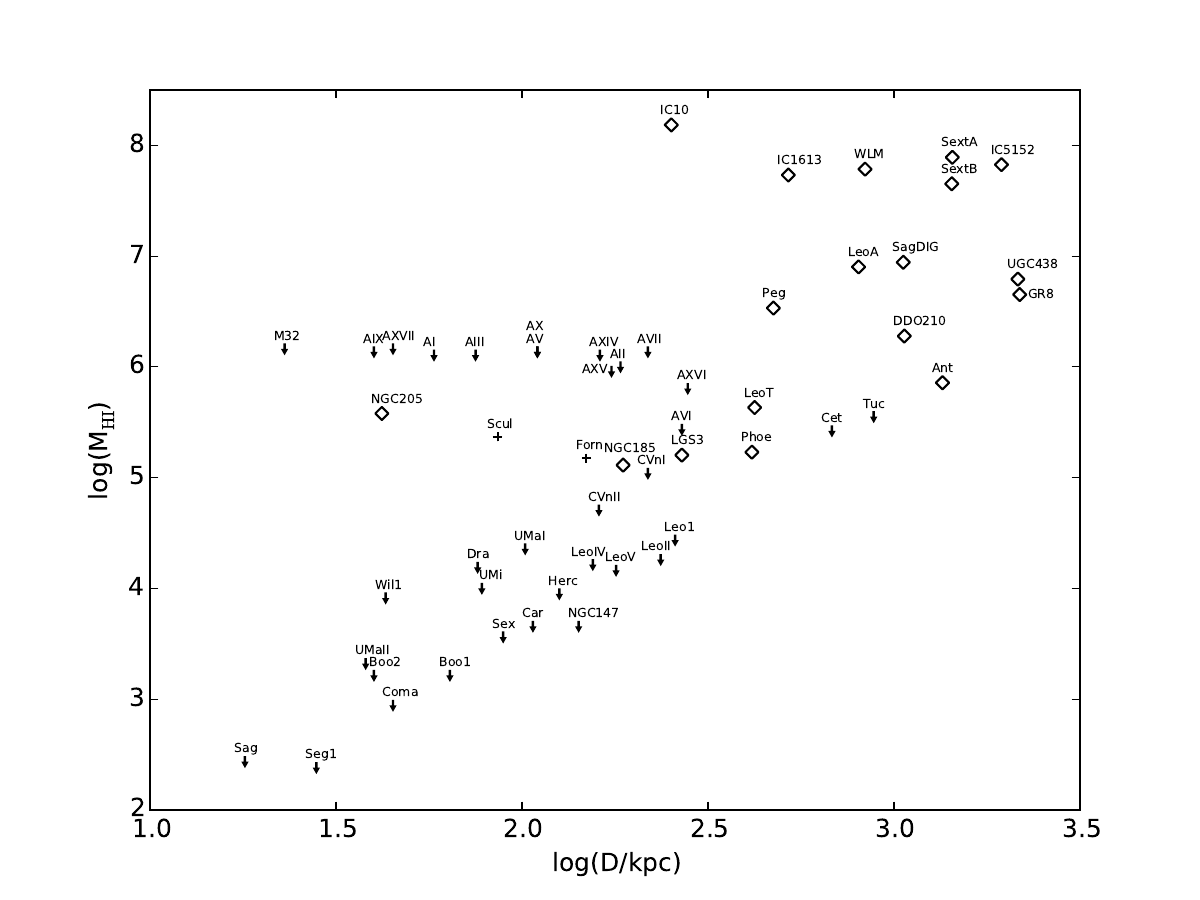}
\caption{HI mass vs. distance to the center of the Milky Way or Andromeda, whichever is closer as tabulated by \cite{mcconnachie12}, for the dwarf galaxies of the Local Group. Downward arrows indicate upper limits, plus signs are ambiguous detections, and diamonds indicate confident detections.}
\label{hidist}
\end{figure*}

Figure \ref{hidist} shows the HI mass or upper limit of each Local Group dwarf vs. the
distance to the center of the Milky Way or Andromeda from the dwarf, whichever is closer to the given satellite as tabulated by \cite{mcconnachie12}. Non-detections are
associated with upper limits in HI mass, and are marked with downward arrows. Confident detections are marked with
diamonds and the two ambiguous detections with plus signs. The apparent lines of upper limits in Figure~\ref{hidist} arise due to the
distance dependence of the HI mass limits.

As illustrated in Figure~\ref{hidist}, there is a cutoff in the distance to the Milky Way or Andromeda within the range of 250 to 280 kpc (which corresponds to log values of 2.4 and 2.45) within which galaxies are largely undetected in HI, and beyond which the majority of galaxies have significant amounts of HI.  A similar type of relationship was noted by other authors \citep{einasto74,lin83,BR00,grebel03}. There are only two satellite galaxies with significant amounts of HI at galactocentric radii less than 250 kpc, NGC 185 and NGC 205.   These galaxies are dwarf ellipticals that have total masses between $10^{8-9}$ M$_{\odot}$, much greater than that of the typical satellites in our sample. The two ambiguous detections, Sculptor and Fornax, are at intermediate Galactocentric distances of 88 and 138 kpc respectively.  Beyond 280 kpc, 14 galaxies are detected confidently at masses greater than $10^5$ M$_{\odot}$, and two are not detected (Tucana and Cetus).   LGS3 and IC10 are additional confident detections that lie between 250-280 kpc.
The mean HI mass of the 16 detected galaxies beyond 250 kpc is 3.1 $\times$ 10$^{7}$ M$_{\odot}$, and the median HI mass is 7.1 $\times$ 10$^{6}$ M$_{\odot}$.

\begin{figure*}
\includegraphics[width=0.98\textwidth]{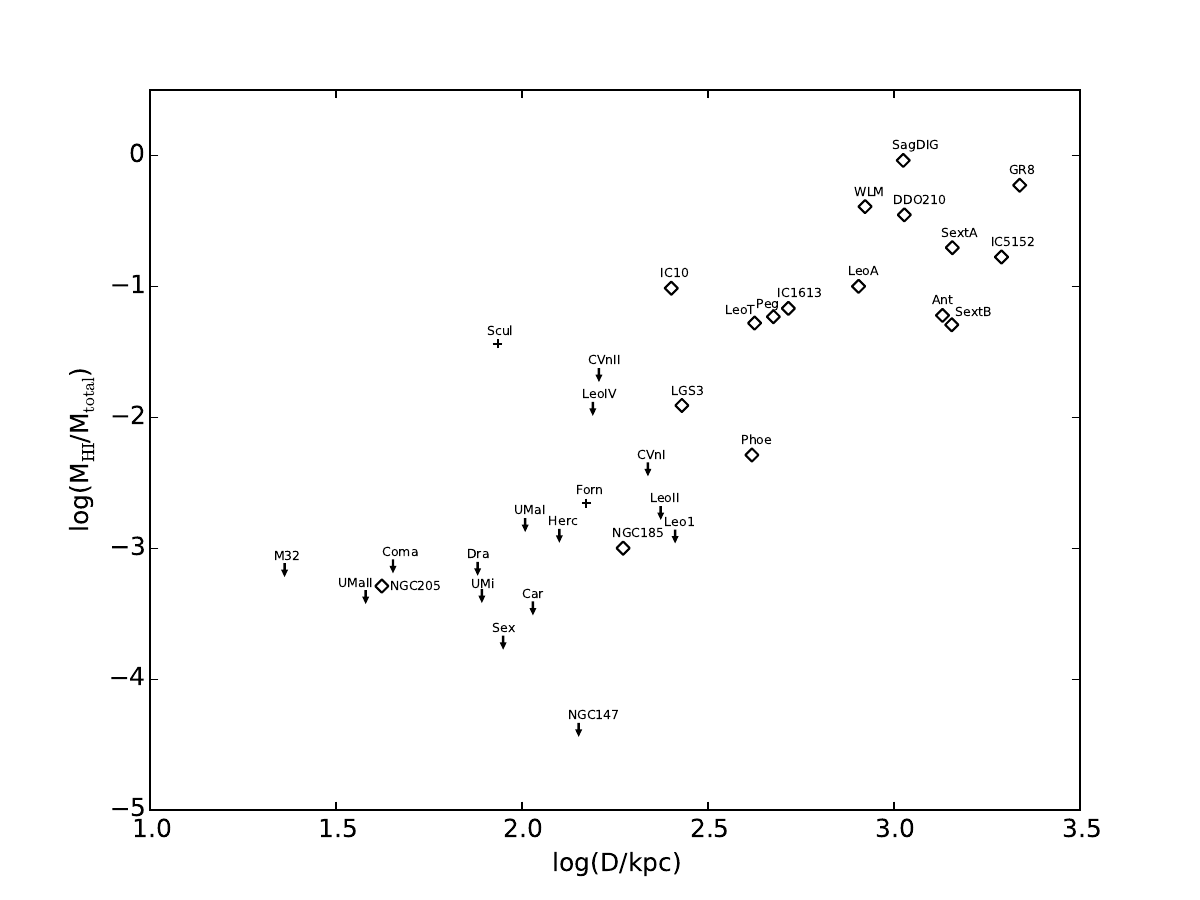}
\caption{HI mass divided by total mass vs. distance to the center of the Milky Way or Andromeda, whichever is closer, for those Local Group dwarf galaxies in Figure~\ref{hidist} with calculated total masses at the time of the original publication.  Symbols are the same as Figure~\ref{hidist} with downward arrows indicating upper limits, plus signs as ambiguous detections, and diamonds as confident detections. Total masses are from Mateo (1998) except Canes Venatici I, Canes Venatici II, Coma Berenices, Hercules, Leo IV, Leo T, Ursa Major I, and Ursa Major II \citep{simon07}, and Leo A \citep{brown07}.}
\label{massdist}
\end{figure*}

\begin{figure*}
\includegraphics[width=0.98\textwidth]{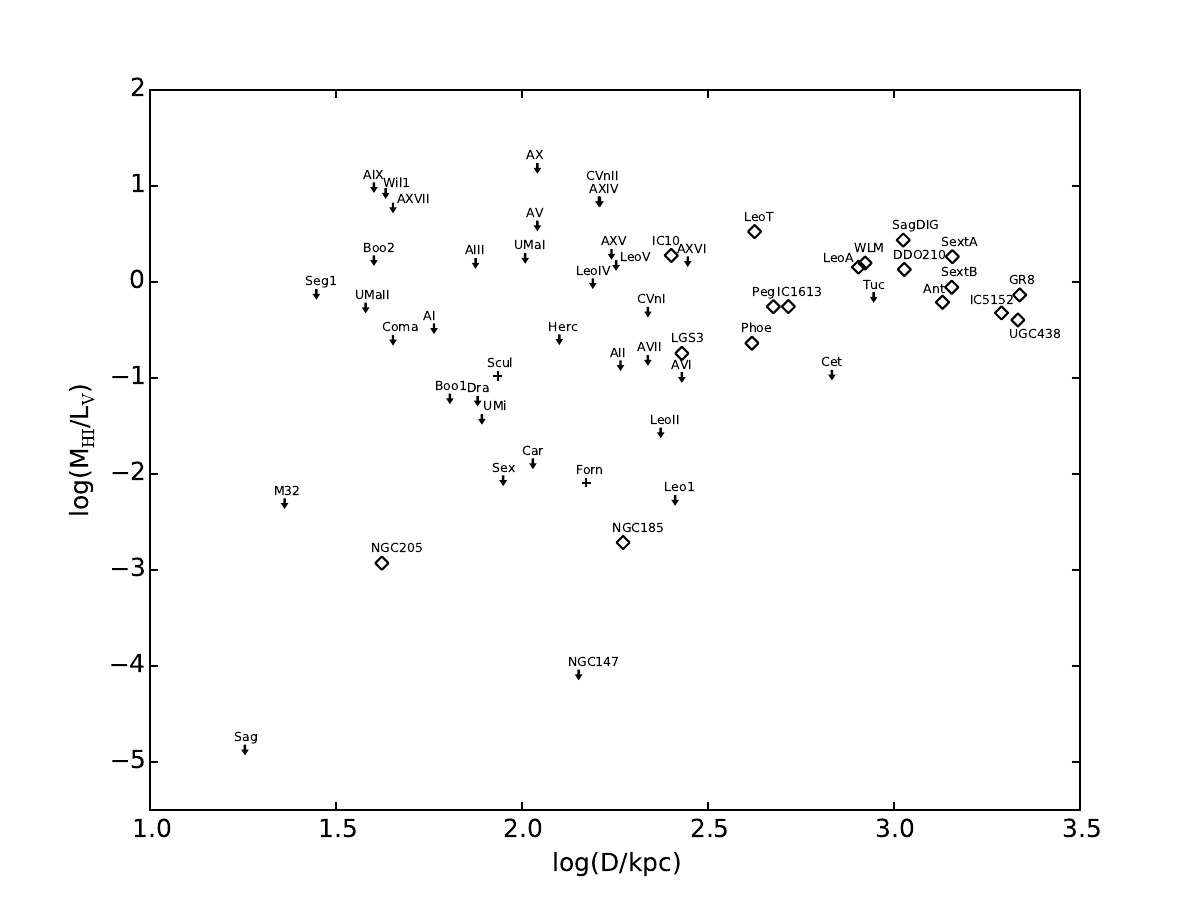}
\caption{HI mass divided by V-band luminosity (in solar units) vs. distance to the center of the Milky Way or Andromeda, whichever is closer for the Local Group dwarf galaxies in Figure~\ref{hidist}. Symbols are the same as Figure~\ref{hidist} with downward arrows indicating upper limits, plus signs as ambiguous detections, and diamonds indicating confident detections. V-band luminosities are from \cite{mcconnachie12}.}
\label{lumdist}
\end{figure*}

Figure~\ref{massdist} shows HI mass normalized by total mass verses galactocentric distance for those dwarfs with measured dynamical masses.  The majority of the dwarf galaxies with measured total masses within 270 kpc have limits on their gas fractions that are less than the gas fractions of galaxies beyond 270 kpc.  We have also plotted the HI mass normalized by V band luminosity (in L$_{\odot}$) in Figure~\ref{lumdist}.  The non-detected dwarf galaxies within 270 kpc have M$_{HI}$/L$_{V}$ limits that are largely comparable to, or lower than, the values of the detections beyond 270 kpc.  Figures~\ref{massdist} and~\ref{lumdist} indicate that our limits on HI mass are significant even accounting for a simple scaling by total mass or luminosity.  

\section{Discussion}
 Including the information for the newly discovered satellites and updating the HI detections of Local Group dwarfs
further supports the idea that those dwarfs at small galactocentric radii have less HI than dwarf galaxies at large radii. We find there is a cutoff at approximately 270 kpc, within which most galaxies are undetected
and beyond which all the galaxies are confident detections with the exception of Cetus and Tucana. The two galaxies with
ambiguous detections, Sculptor and Fornax, are at distances of 88 and 138 kpc. These galaxies could be devoid of gas
and the detections are the result of chance superpositions, or the nearby clouds may have originated from the dwarf and been offset from the optical center during the process of gas removal. The lack of HI in most satellites within about 270 kpc of the Galactic and M31 center indicates the dominant gas removal mechanism is related to the proximity to the primary galaxy.  It is important
to note that many of the newly discovered SDSS dwarfs are near the SDSS detection limit, and if similar objects existed at greater
distances they would not have been discovered \citep{koposov07b}. If these hypothetical dwarfs were deficient in HI, the lower right region of Figure~\ref{hidist} as it stands now would be underpopulated. The existence of such galaxies would not affect the conclusion that those satellites within 270 kpc tend to be HI deficient.

These galaxies are a potential source of star formation fuel if their gas is accreted by the Milky Way. If we examine the case of the Milky Way alone, we can estimate the contributed amount of HI gas. First we assume that each of the satellite galaxies within 270 kpc had an average of 10\% of their measured dynamical mass in neutral hydrogen, and that the gas of the galaxy was completely integrated into the Milky Way's disk. Galaxies or streams\footnote{This calculation includes destroyed galaxies seen as streams, including the Monoceros Stream, the Orphan Stream, and the Virgo overdensity \citep{newberg02, belokurov07a, newberg07}. The number of streams which have fully integrated into the disk is unknown and their contribution is not included.} with unknown masses are assumed to have the same HI mass as Leo T (4.3 $\times$ 10$^{5}$ M$_{\odot}$) on average. In this scenario, the total amount of gas recently accreted by the Milky Way from known satellite dwarf galaxies would be $\sim$ 2.1 $\times$ 10$^{7}$ M$_{\odot}$. 

The previous calculation does not account for incompleteness in the total satellite galaxy count due to limited sky coverage and survey detection limits. Tollerud et al. (2008\nocite{tollerud08}) predicted the total number of Milky Way satellite galaxies within various radii assuming that the Via Lactea N-body simulation \citep{diemand07} is a good representation of the spatial distribution of Milky Way satellites and also by using the SDSS detection efficiencies given in Koposov et al. (2007\nocite{koposov07a}). They estimate there are 322 satellite galaxies within 300 kpc of the Milky Way, with a 98\% confidence range between 246 and 466 satellites. These galaxies must be relatively faint, or they would have otherwise been discovered, so it is reasonable to say that on average their HI mass may be similar to that of Leo T. Adding in the contribution from the predicted but undiscovered satellite galaxies, the total amount of HI recently contributed to the Milky Way by dwarfs is in the range of 1.2 - 2.1 $\times$ 10$^{8}$ M$_{\odot}$. Since most chemical evolution models suggest we need an average of  $\sim$1 M$_{\odot}$ yr$^{-1}$ of infalling fuel over the past 5-7 Gyrs in order to explain the metallicity of the long lived G and K stars \citep{chiappini01, fenner03}, dwarf galaxies alone cannot provide sufficient fuel to the Milky Way in the current era.

\subsection{Dwarf Gas Loss}

Proposed methods by which dwarf satellites have their gas removed include ram pressure stripping, tidal stripping, feedback from
supernovae or stellar winds, and the effects of reionization. In the process known as ram pressure stripping, as a satellite moves through the halo medium it experiences a pressure whose strength
depends on the satellite's velocity, total mass, and gas density and the properties of the ambient gas \citep{gunn72}. If the orbit of a dwarf brings
it into a region of sufficient density the pressure will be great enough to allow the gas to escape the potential well of 
the satellite. If ram pressure stripping is taking place, we can
estimate the density of the diffuse hot halo gas that a satellite has experienced. The general equation describing the condition necessary for stripping to take place is,
\begin{equation}
n_{halo} \sim \frac{\sigma^{2}\,n_{gas}}{3\,v_{sat}^{2}} \; cm^{-3}
\label{1}
\end{equation}
where $n_{halo}$ is the ambient gas number density, $\sigma$ is the central stellar velocity dispersion of the dwarf, $v_{sat}$ is
the relative motion of the dwarf through the medium, and $n_{gas}$ is the average gas density of the dwarf in the inner regions. It should be noted that this equation assumes that stripping is instantaneous, occurs in a homogeneous medium, and does not trigger star formation which can heat the gas and increase stripping efficiency; some of these factors may play an important role (e.g., Mayer et al. 2006\nocite{mayer06}).

Another way to strip a galaxy of its gas is via the effects of massive star evolution. Internal mechanisms such as stellar winds and
supernovae may cause gas loss from the shallow potential wells of the dwarfs. 
Though star formation and the resulting feedback may play a role in heating the gas and making it easier to strip, it is unlikely to result in the distant dependent mass loss shown in Figure~\ref{hidist}.  This is emphasized by Figures~\ref{massdist} and \ref{lumdist}, which show the limits on the gas content of the dwarfs is significant even when scaling by total mass and stellar content.  If gas-loss due to stellar feedback was dominant, a relationship between gas content and the quantity of stars and/or the total mass of the galaxy (depth of the potential well) may be apparent.  Strigari et. al. (2008\nocite{strigari08}) have shown that both the newly discovered and previously known dwarf spheroidals have a similar total mass of $\sim$10$^{7}$ M$_{\odot}$ interior to 300 pc.  Since the gas would not escape more easily from the nearby dwarfs, there is no evidence that stellar feedback is the dominant gas loss mechanism. Though the products of stellar evolution can potentially also contribute HI to galaxies \citep{vanloon06,bouchard05}, it would not create the distance dependent effect seen in Figures ~\ref{hidist} - \ref{lumdist}.  

An additional gas loss mechanism is photoionization during the epoch of reionization \citep{gnedin06, dijkstra04, madau08, ricotti05}.  The effects of reionization inhibit the ability of the lowest mass halos to accrete gas.  It has been proposed that the smallest dwarf galaxies of the Local Group formed their stars before reionization when they were still capable of accreting gas \citep{gnedin06}.  The mass scale for the halos that are able to accrete gas from the IGM after reionization is somewhat uncertain \citep{dijkstra04}, however the gas-rich, low mass Leo T is difficult to explain unless its dark matter halo extends out to much larger radii than the observed baryons. In the case of reionization it may be possible that galaxies further from the main source of ionization would be more likely to retain gas.   If the Milky Way and Andromeda were major sources of ionization at early times and there is a correlation with the current and past galactocentric distances of the dwarfs, then reionization may play a role in the HI-distance trend.  Given the amount of time for the Local Group galaxies to evolve since reionization (e.g., Moore et al. 2006\nocite{moore06}), it seems unlikely that reionization could lead to the present day HI-distance effect. 

The lack of significant HI in nearby dwarfs is due to a distant dependent mechanism. The two most widely studied distance dependent gas-loss mechanisms are tidal and ram pressure stripping, with simulations showing that the combination of tides and ram pressure is more effective than either mechanism alone.  Ram pressure stripping is the dominant gas loss mechanism in the simulations \citep{mayer06}, with tides enhancing the effectiveness of ram pressure stripping by lowering the depth of the satellite's potential well.  
The limited effect of tidal forces in stripping the gas from the satellites is evident from calculations of tidal radii of the dwarf galaxies and studies of the galaxies stellar components.  The tidal radius for a 10$^{7}$ M$_{\odot}$ dwarf galaxy at a distance of 20 kpc from the center of the Milky Way is on the order of 1 kpc, and increasing the total mass of the satellite galaxy only serves to increase the tidal radius \citep{battaner00}.  This is significantly larger than the extent of HI in Leo T and is three times the average stellar extent of the newly discovered dwarf galaxies \citep{strigari08}. The stellar component of 18 Local Group dwarf galaxies was examined by Strigari et al. (2008\nocite{strigari08}) to search for the current effects of tidal forces on individual dwarfs.  They searched for gradients in the line of sight stellar velocities across the face of the galaxies (including Willman I, Coma Berenices, and Ursa Major II, which are all within 44 kpc of the Milky Way) and found no significant detection of streaming motions indicative of tidal disruption. The dominant gas loss mechanism is likely to be ram pressure stripping for the dwarf satellites in the Local Group, as also concluded by BR00, and we address this further below.  

\subsection{Halo Density}  
Leo T still has a significant amount of HI and does not appear to have been affected by ram pressure stripping or tidal disruption \citep{ryanweber08,strigari08}. Given the total mass of Leo T and its diffuse stellar component, it is likely similar to the progenitors of the newly discovered dwarfs which do not have HI. The diffuse halo component required to completely strip this type of galaxy can be calculated. We assume a Leo T-like value for $\sigma$ of $7.5$ km s$^{-1}$ \citep{simon07}. We also estimate a range of possible dwarf gas densities; for the low end of the range, we take the mean gas density in the central region of Leo T, n$_{gas} =$0.12 cm$^{-3}$, and for the high limit we take the central density calculated from fitting a Plummer model to the column density profiles in \cite{ryanweber08}, which yields $n_{gas} =$0.44 cm$^{-3}$. We approximate the value of $v_{sat}$ as the one dimensional velocity dispersion of 60 km s$^{-1}$ for
Local Group dwarfs \citep{vandenbergh99}. 
Using these values and assuming the dwarfs are on circular orbits or experiencing their initial infall, we find that the newly discovered, gas-free dwarfs likely experienced a halo density greater than $n_{halo} \sim$ 0.6-2.3 
$\times$ 10$^{-3}$ cm$^{-3}$ at the distance limit where the dwarfs have HI, or $\sim$270 kpc. Observations indicate densities on the order of 10$^{-3}$ cm$^{-3}$ are extremely unlikely at this distance \citep{gaensler08, sembach03, putman04, peek07}, as do simulations and calculations of the hot halo density profile (Kaufmann et al. 2008, 2007; Sommer-Larsen 2006; Maller \& Bullock 2004\nocite{kaufmann07,kaufmann08,sommerlarsen06, maller04}). Figure~\ref{halo} shows several of the hot halo density profiles from the simulations drop towards $10^{-5}$ cm$^{-3}$ at distances greater than $200$ kpc.

The most likely solution to the above is that the dwarf orbits have brought them closer to the center of the parent galaxy than we see them today, and they therefore experienced a much higher halo density than that present in their current environment. In addition, they would have attained a much greater velocity through the halo medium as they approached the galaxy, and because the required density for stripping scales as $v^{-2}$, they would require lower densities for stripping during this portion of their orbit. It is also possible they have traveled through their parent galaxy's disk before arriving at their current position. Eccentric orbits could cause an overestimate of the typical stripping radius based on Figure~\ref{hidist}, since the satellites would spend more time at apogalacticon than perigalacticon.

Several dwarfs do have proper motion measurements which give an estimation of their space velocity and orbital characteristics. 
The velocity at perigalacticon can be calculated by using the current radius and velocity in the galaxy rest frame to find the specific angular momentum of the orbit, $J = vR$. Using angular momentum conservation, the velocity at perigalacticon is $v_{peri} = J/R_{peri}$. 
We calculated $v_{peri}$ for the dwarfs with proper motions and orbital analyses, Carina \citep{piatek03}, Ursa Minor \citep{piatek05}, Sculptor \citep{piatek06}, and Fornax \citep{piatek02, walker08}.  
We assume typical values for the central stellar velocity dispersion, $\sigma$ =  10 \kms \citep{mateo98}, and a central density, $n_{gas}$ = 0.5 cm$^{-3}$.
For all of the dwarfs with proper motions there is a range of possible orbits and perigalacticons, and thus a range in the densities required for stripping. The 90\% confidence range for the distance at perigalacticon from the proper motion references as well as the maximum and minimum density required for stripping within that range is listed for each dwarf in Table 4. In the case of Carina, we calculate a lower limit for the halo density of 8.5 $\times$ 10$^{-5}$ cm$^{-3}$ at the most likely perigalacticon of 20 kpc. For Ursa Minor, the calculated lower limit for the density of the halo at the most likely perigalacticon of 40 kpc is 2.1 $\times$ 10$^{-4}$ cm$^{-3}$. Using the Sculptor orbital characteristics we calculate a halo density of 2.7 $\times$ 10$^{-4}$ cm$^{-3}$ at the most likely perigalacticon of 68 kpc.  The case of Fornax is interesting because at 137 kpc, it may be near perigalacticon \citep{piatek02}.  The most likely perigalacticon for Fornax is 118 kpc and at that distance the required density for stripping is $\sim$3.1 $\times$ 10$^{-4}$ cm$^{-3}$. 
We note that these calculations of the halo density do not include tidal effects which may play a small role in contributing to the effectiveness of ram pressure stripping for the closest perigalactica (as previously discussed).

Sculptor and Fornax are the most distant dwarf galaxies in Table~\ref{Table 4} and are also labeled as ambiguous detections indicating there are clouds in the vicinity of these galaxies which may or may not be associated (see Figures~\ref{ambig} and~\ref{hidist}).  Since in
both cases the HI clouds are offset from the optical center of the dwarf, even if the gas
is associated with the dwarf galaxies it appears to have been partially removed.
The halo density estimate holds for these two galaxies if the gas has been recently stripped or is not associated with the dwarf.
 
We can now come back to the newly discovered dwarf galaxies (with a Leo T like progenitor) and give them a velocity at perigalacticon that is closer to that obtained by the above dwarf galaxies to estimate a more likely density required to strip them.  We do not know the actual perigalactica of these galaxies, but if they were moving at velocities between 200 - 400 \kms ~halo densities of  $\sim 1.0-4.2 \times 10^{-4}$ cm$^{-3}$ would be required to strip them.  This is much closer to the expected halo densities at the current distance of many of the newly discovered dwarf galaxies.
 
\subsection{Comparison to Halo Models}
\begin{figure*}
 \includegraphics[width=5in,angle=90]{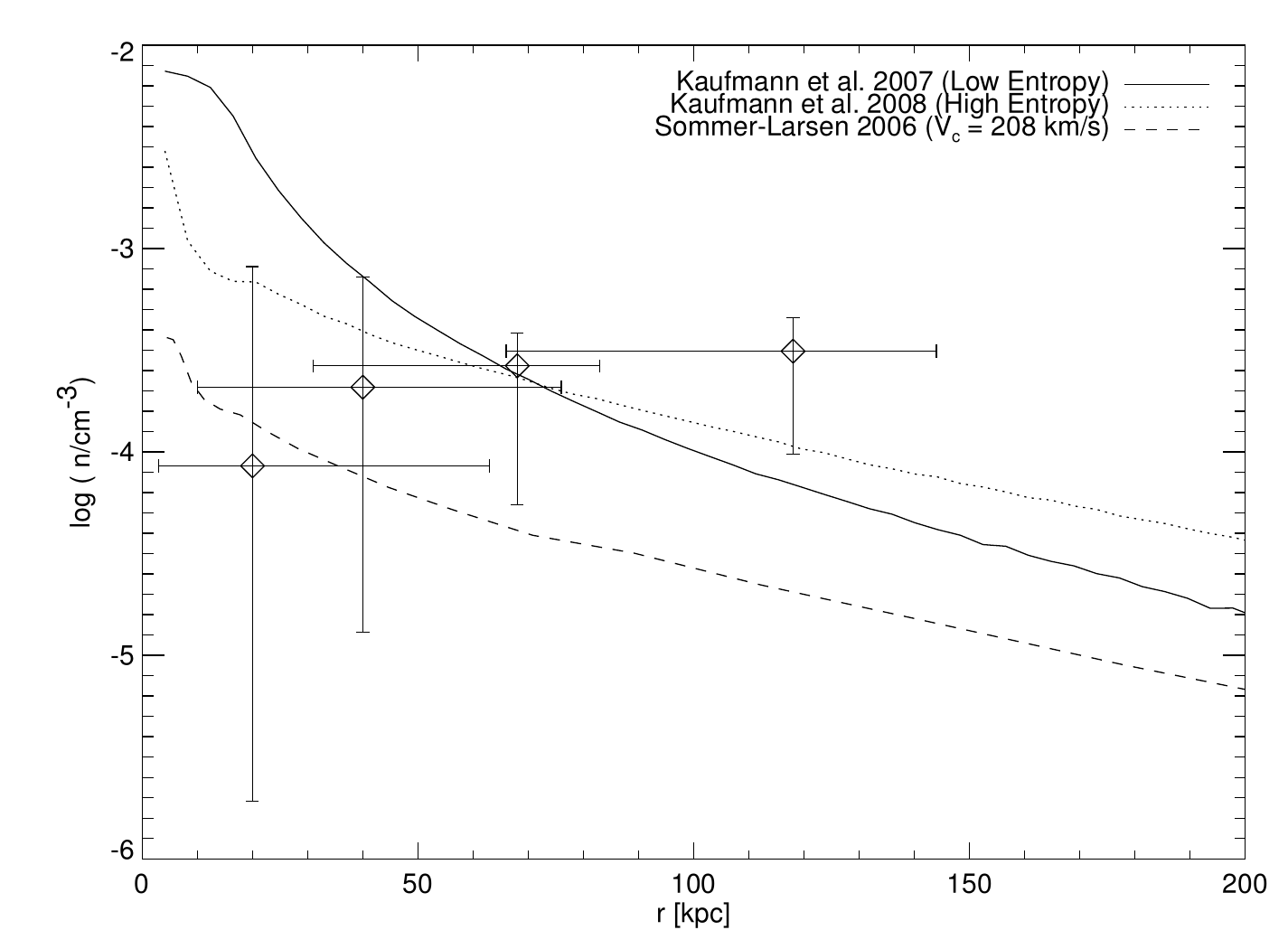}
\caption{The diamonds represent estimates for the halo density as a function of Galactocentric distance calculated from the orbital characteristics of Carina, Ursa Minor, Sculptor, and Fornax (from left to right).  The lines represent the model halo density profiles with the references given in the legend of the plot.}
\label{halo}
\end{figure*}

The densities of the halo derived from the dwarf galaxies with proper motion estimates can be compared to theoretical models of hot gas confined within a Milky Way-sized dark matter halo. Figure~\ref{halo} shows halo density vs. Galactocentric distance with the range in distances and halo densities for the dwarf galaxies taken from Table 4.  The solid line is the theoretical density profile for gas whose initial distribution traces the central cusp in the NFW halo (referred to as the low entropy model), while the dotted line represents the density profile which results from an initial gas distribution with a central core of high entropy \citep{kaufmann07,kaufmann08}. The high entropy model is expected for haloes which have experienced pre-heating feedback early in their histories, and implies a more extended distribution for the hot halo gas, as well as an extended cloud population \citep{rasmussen06, li07}. The remaining density model plotted on Figure~\ref{halo} is from \cite{sommerlarsen06} and is from high resolution cosmological SPH simulations of a Milky-Way like galaxy in a $\Lambda$CDM cosmology. The densities derived from the stripping of the dwarf galaxies are broadly consistent with the theoretical profiles.  The exception is the value calculated with Fornax which predicts a halo density that is higher than the models in most cases.   If the gas clouds in the vicinity of Fornax are in the process of being stripped, an overestimate could be due to the calculation of the halo density being for the complete stripping of the gas from the dwarf.

\begin{figure*}
\includegraphics[width=5in,angle=90]{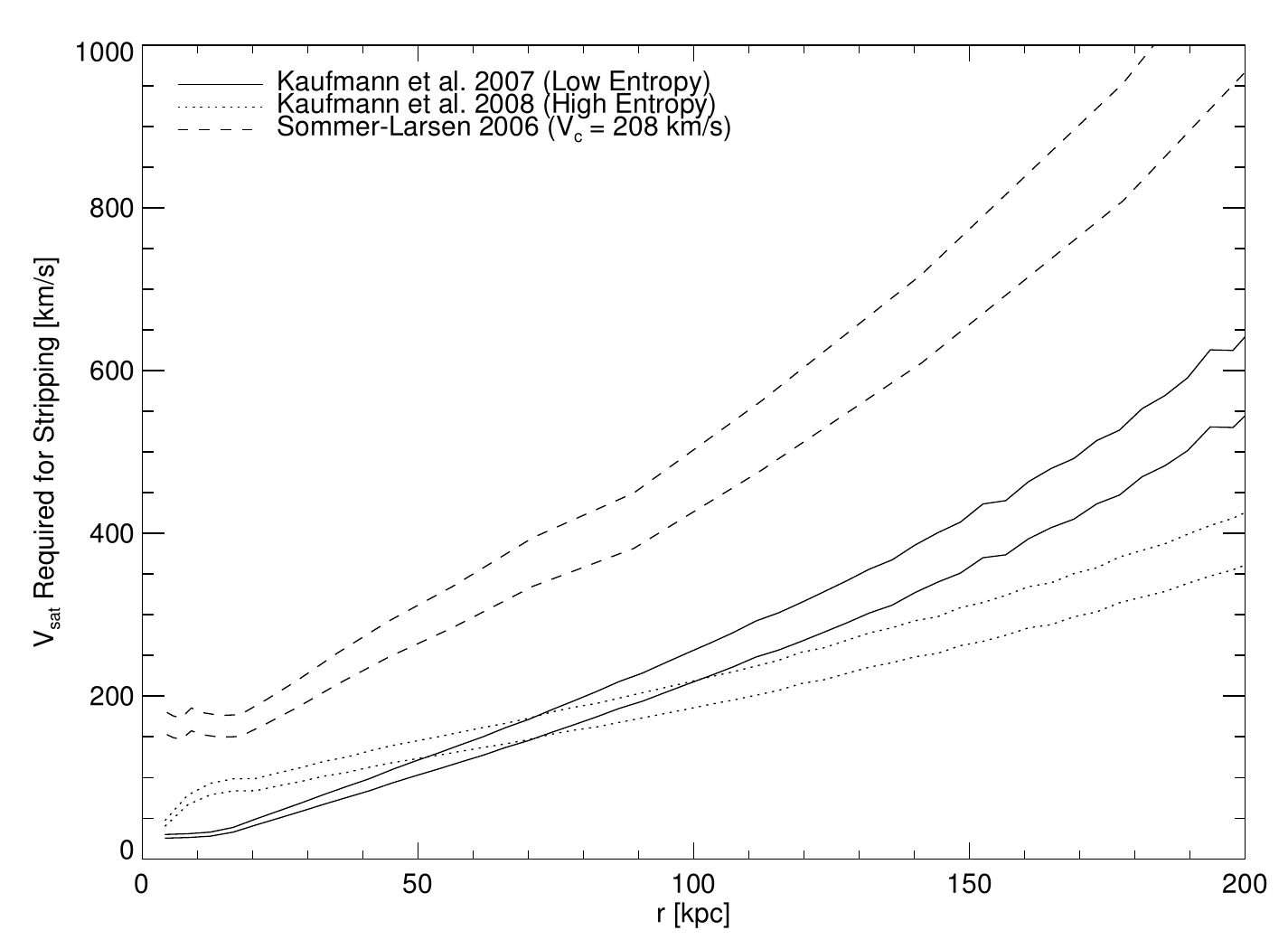}
\caption{A plot of the satellite velocities required for ram-pressure stripping of a satellite galaxy using the density profiles of the Milky Way's hot halo gas shown in Figure~\ref{halo}.  The two lines represent the range of typical satellite galaxy characteristics: n$_{gas} \sim0.1-0.8$ cm$^{-3}$, $\sigma \sim 5-10$ \kms.}
\label{halovel}
\end{figure*}

It is possible, given a halo model and a typical range of dwarf galaxy characteristics, to calculate the velocity required to strip a satellite galaxy at a given radius. This is plotted in Figure~\ref{halovel} for the three halo density models previously discussed and using a range in dwarf galaxy velocity dispersions ($\sigma = 5-12$ \kms) and central densities (n$_{gas} = 0.1-0.8$ cm$^{-3}$). As additional proper motions are determined this plot can be used to check consistency between the ram pressure stripping scenario and the halo density models. Also, if there is independent evidence that a dwarf is being stripped at its current radius (e.g., head-tail structure; Quilis \& Moore 2001\nocite{quilis01}), Figure~\ref{halovel} could be used to estimate a lower limit on the velocity of the satellite. 

\section{Summary} We conducted an analysis of the HI content of Local Group dwarfs including those extremely low mass
dwarfs discovered through November 2008 via SDSS and deep surveys of the M31 environment. We used HIPASS and LAB data to determine the HI mass or upper limits on the SDSS
dwarfs and have made several conclusions.
\begin{itemize}
\item All of the Milky Way SDSS dwarfs except Leo T are devoid of gas to the level of our detection limits (see Table 1).  The upper
limits are under 10$^{5}$ M$_{\odot}$ (except for one at $1.2 \times 10^5$ M$_{\odot}$), and all limits are lower than the HI mass of any known dwarf galaxy with an HI detection.  The newly discovered Andromeda dwarfs also appear to be devoid of gas, but the limits are higher (M$_{HI} < 10^{6.2}$~\Msun).  This result is consistent with the lack of recent
star formation in these galaxies.
\item  Local Group dwarf galaxies at small galactocentric distances (\lessapprox 270 kpc) tend to not have HI while those at
larger galactocentric distances usually do with HI masses above 10$^{5}$ M$_{\odot}$.   
The exceptions at small galactocentric distances are the two higher total mass dE's and two ambiguous detections (Fornax and Sculptor) at 88 and 138 kpc from the Milky Way for which the clouds detected may or may not be associated with the dwarf galaxies.  
There is a clear relationship between galactocentric distance and HI content for dwarf galaxies in the Local Group.  This relationship is still significant when scaling the HI mass by the total mass or luminosity of the dwarf galaxy.
\item By assuming ram pressure stripping is the dominant gas loss mechanism and taking typical characteristics of the dwarf galaxies with gas, we approximate the density of the Galactic halo necessary to strip the Local Group dwarf galaxies.  For those dwarfs with proper motions, we calculate the most likely velocity at perigalacticon and determine limits or approximate values of the Galactic halo density at specific distances from the center of the Galaxy. 
This method estimates the Milky Way's halo density as greater than $\sim$ 8.5 $\times$ 10$^{-5}$, 2.1 $\times$ 10$^{-4}$, 2.7 $\times$ 10$^{-4}$, and 3.1 $\times$ 10$^{-4}$ cm$^{-3}$ at 20, 40, 68, and 118 kpc respectively in order to strip the galaxies.  These values are generally consistent with theoretical models of the hot gas within the Milky Way's extended halo.   We also calculate the velocities required to strip dwarf galaxies without known proper motions given these theoretical halo gas profiles. 
\item  Assuming that the HI gas was stripped and integrated into the Milky Way's disk, and that the satellite galaxy progenitors had typical galaxy characteristics, we estimate that accretion of gas from known stripped galaxies and streams would have provided $\sim 1.1 \times 10^{7}$ M$_{\odot}$ of HI gas to the Milky Way. If the incompleteness in the satellite galaxy count is corrected, we expect about $\sim 1.2-2.1 \times 10^{8}$ M$_{\odot}$ of HI mass to be accreted by the Milky Way. This is not enough to sustain the star formation of the Milky Way in the current era.
\end{itemize}
Thanks to Fabian Heitsch, Mario Mateo, Joshua E. G. Peek, Oleg Gnedin, Marla Geha, Erik Tollerud, Joe Wolf, Jacco van Loon, Tobias Westmeier, Beth Willman, and Kristine Spekkens for helpful discussions. We would also like to thank Kevin Douglas for providing code which helped with the analysis of the LAB data, Tobias Kaufmann and Jesper Sommer-Larsen for providing the hot halo density profile, Joshua E. G. Peek and Kevin Douglas for reducing the GALFA data, and the Research Corporation for partial funding. We would also like to acknowledge the helpful comments of the anonymous referee. This research has made use of the NASA/IPAC Extragalactic Database (NED) which is operated by the Jet Propulsion Laboratory, California Institute of Technology, under contract with the National Aeronautics and Space Administration.  The Parkes telescope is part of the Australia Telescope which is funded by the Commonwealth of Australia for operation as a National Facility managed by CSIRO.
\clearpage
\bibliographystyle{apj}
\bibliography{msnew}

\clearpage
\tabletypesize{\scriptsize}
\begin{deluxetable*}{llllccclc}
\tabletypesize{\scriptsize}
\tablewidth{0pt} 
\tablecaption{HI Mass of Newly Discovered Satellites}  
\tablehead{\colhead{Object}  &\colhead{Data} &\colhead{$\alpha$}& \colhead{$\delta$} & \colhead{Optical
Velocity}&\colhead{Galactic Interference} & \colhead{Distance} & \colhead{HI Mass} & \colhead{References}\\
\colhead{}&\colhead{}&\multicolumn{2}{c}{(J2000)}&\colhead{(km s$^{-1}$)}&\colhead{Range (km
s$^{-1}$)}&\colhead{(kpc)}&\colhead{$(M_\odot)$}&\colhead{}}
\startdata 
Ursa Major II&LAB&$08^{h}51^{m}30^{s}$&$+63^{\circ}07^{\prime}48^{\prime\prime}$&-117&-70 to 30&30&$<2.3\times 10^{3}$&a,e\\
Leo T&HIPASS&$09^{h}34^{m}53^{s}$&$+17^{\circ}02^{\prime}52^{\prime\prime}$&38&-73 to 46&420&$\sim 4.3 \times 10^{5}$&b,e\\
Segue I&HIPASS&$10^{h}07^{m}04^{s}$&$+16^{\circ}04^{\prime}56^{\prime\prime}$&206&-73 to 46&23&$< 265$&c,k\\
Ursa Major I&LAB&$10^{h}34^{m}53^{s}$&$+51^{\circ}55^{\prime}12^{\prime\prime}$&-55&-70 to 25&100&$< 2.5 \times 10^{4}$&d,c,e\\
Willman I&LAB&$10^{h}49^{m}22^{s}$&$+51^{\circ}03^{\prime}04^{\prime\prime}$&-12&-75 to 5&38&$<9\times10^{3}$&i,l\\
Leo V&HIPASS&$11^{h}31^{m}09^{s}$&$+02^{\circ}13^{\prime}12^{\prime\prime}$&173&-72 to 72&180&$<1.6\times 10^{4}$&j\\
Leo IV\tablenotemark{a}&HIPASS&$11^{h}32^{m}57^{s}$&$-00^{\circ}32^{\prime}00^{\prime\prime}$&132&-60 to 60&160&$< 1.8\times 10^{4}$&c,e\\
Coma Berenices&HIPASS&$12^{h}26^{m}59^{s}$&$+23^{\circ}54^{\prime}15^{\prime\prime}$&98&-60 to 33&44&$< 968$&c,e\\
Canis Venetici II&LAB&$12^{h}57^{m}10^{s}$&$+34^{\circ}19^{\prime}15^{\prime\prime}$&-129&-50 to 25&150&$<5.6\times
10^{4}$&c,e\\
Canis Venetici I&LAB&$13^{h}28^{m}04^{s}$&$+33^{\circ}33^{\prime}21^{\prime\prime}$&31&-20 to 25&220&$<1.2\times
10^{5}$&c,e,f\\
Bo\"{o}tes II&HIPASS&$13^{h}58^{m}00^{s}$&$+12^{\circ}51^{\prime}00^{\prime\prime}$&-117&-33 to 33&60&$<1.8\times 10^{3}$&h,m\\
Bo\"{o}tes I&HIPASS&$14^{h}00^{m}06^{s}$&$+14^{\circ}30^{\prime}00^{\prime\prime}$&96&-33 to 33&60&$<1.8\times 10^{3}$&g,l\\
Hercules&HIPASS&$16^{h}31^{m}02^{s}$&$+12^{\circ}47^{\prime}30^{\prime\prime}$&45&-60 to 60&140&$<9.8\times 10^{3}$&c,e\\
\enddata
\label{Table 1}
\tablecomments{Upper limits for undetected objects are the  HIPASS or LAB $5\sigma$ detection limits at the distance of the dwarf. References for the optical data: a: \cite{zucker06b}, b: \cite{irwin07}, c: \cite{belokurov07b}, d: \cite{willman05b}, e: \cite{simon07}, f: \cite{zucker06a}, g: \cite{belokurov06b}, h: \cite{walsh07}, i: \cite{willman05b}, j: \cite{belokurov08}, k: \cite{geha09}, l: \cite{martin07}, m: \cite{koch09}}
\tablenotetext{a}{\scriptsize Limit is higher for this galaxy due to confusion with a HVC complex that lies at a similar position and velocity.}
\end{deluxetable*}

 \tabletypesize{\scriptsize}
 \begin{deluxetable*}{lllccll}
\tablewidth{0pt} 
\tablecaption{HI Mass of Andromeda Satellites}  
\tablehead{\colhead{Object}  &\colhead{$\alpha$}& \colhead{$\delta$} &\colhead{D$_{\odot}$}& \colhead{D$_{M31}$} &
\colhead{HI Mass}&\colhead{References}\\
\colhead{}&\multicolumn{2}{c}{(J2000)}&\colhead{(kpc)}&\colhead{(kpc)}&\colhead{$(10^{6} M_\odot)$}&\colhead{}}
\startdata 
And I&$00^{h}45^{m}40^{s}$&$+38^{\circ}02^{\prime}28^{\prime\prime}$&745&59&$< 1.4$&a,b,c\\
And II&$01^{h}16^{m}30^{s}$&$+33^{\circ}25^{\prime}09^{\prime\prime}$&652&185&$< 1.1$&a,b,c\\
And III&$00^{h}35^{m}34^{s}$&$+36^{\circ}29^{\prime}52^{\prime\prime}$&749&76&$< 1.4$&a,b,c\\
And V&$01^{h}10^{m}17^{s}$&$+47^{\circ}37^{\prime}41^{\prime\prime}$&774&110&$< 1.5$&b,d\\
And VI&$23^{h}51^{m}46^{s}$&$+24^{\circ}34^{\prime}57^{\prime\prime}$&783&269&$< 0.3$&a,b,c,e\\
And VII&$23^{h}26^{m}31^{s}$&$+50^{\circ}41^{\prime}31^{\prime\prime}$&763&219&$< 1.5$&b,d,f\\
And IX&$00^{h}52^{m}53^{s}$&$+43^{\circ}11^{\prime}45^{\prime\prime}$&765&42&$< 1.5$&b,g\\
And X&$01^{h}06^{m}34^{s}$&$+44^{\circ}48^{\prime}16^{\prime\prime}$&783&112&$< 1.5$&b,h\\
And XI&$00^{h}46^{m}20^{s}$&$+33^{\circ}48^{\prime}05^{\prime\prime}$&\nodata&103&$< 1.5$&i\\
And XII&$00^{h}47^{m}27^{s}$&$+34^{\circ}22^{\prime}29^{\prime\prime}$&\nodata&95&$< 1.5$&i\\
And XIII&$00^{h}51^{m}51^{s}$&$+33^{\circ}00^{\prime}16^{\prime\prime}$&\nodata&116&$<1.5$&i\\
And XIV&$00^{h}51^{m}35^{s}$&$+29^{\circ}41^{\prime}49^{\prime\prime}$&740&167&$<1.4$&k\\
And XV&$01^{h}14^{m}19^{s}$&$+38^{\circ}07^{\prime}03^{\prime\prime}$&630&170&$<1.0$&j\\
And XVI&$00^{h}59^{m}30^{s}$&$+32^{\circ}22^{\prime}36^{\prime\prime}$&525&270&$<0.7$&j\\
And XVII&$00^{h}37^{m}07^{s}$&$+44^{\circ}19^{\prime}20^{\prime\prime}$&794&45&$<1.6$&l\\
M32&$00^{h}42^{m}42^{s}$&$+40^{\circ}51^{\prime}54^{\prime\prime}$&805&22&$<1.6$&a\\
NGC 147&$00^{h}33^{m}12^{s}$&$+48^{\circ}20^{\prime}12^{\prime\prime}$&725&115&$<0.005$&a\\
NGC 185&$00^{h}38^{m}58^{s}$&$+48^{\circ}20^{\prime}12^{\prime\prime}$&620&185&0.13&a\\
NGC 205&$00^{h}40^{m}22^{s}$&$+41^{\circ}41^{\prime}24^{\prime\prime}$&815&32&0.38&a\\
\enddata
\label{Table 2}
\tablecomments{a: \cite{mateo98}, b: \cite{vandenbergh06}, c: \cite{vandenbergh72}, d: \cite{armandroff98}, e: \cite{armandroff99}, f: \cite{karachentsev99}, g: \cite{zucker04}, h: \cite{zucker07}, i: \cite{martin06}, j: \cite{ibata07}, k: \cite{majewski07}, l: \cite{irwin08}}
\end{deluxetable*}
 \begin{deluxetable*}{lllccll}
 \tabletypesize{\scriptsize}
\tablewidth{0pc} 
\tablecaption{HI Mass of Additional Local Group Satellite Galaxies}  
\tablehead{\colhead{Object} &\colhead{$\alpha$}& \colhead{$\delta$} & \colhead{Optical Velocity} &
\colhead{D$_{\odot}$} & \colhead{HI Mass}&\colhead{References}\\
\colhead{}&\multicolumn{2}{c}{(J2000)}&\colhead{(km s$^{-1}$)}&\colhead{(kpc)}&\colhead{$(10^{6}$ M$_\odot)$}&\colhead{}}
\startdata 
WLM&$00^{h}01^{m}58^{s}$&$-15^{\circ}27^{\prime}48^{\prime\prime}$&-78&925&61&a,j\\
IC 10&$00^{h}20^{m}25^{s}$&$+59^{\circ}17^{\prime}30^{\prime\prime}$&-344&825&153&a\\
Cetus&$00^{h}26^{m}11^{s}$&$-11^{\circ}02^{\prime}40^{\prime\prime}$&-87&755&$<0.29$&f,aa\\
SMC\tablenotemark{a}&$00^{h}52^{m}44^{s}$&$-72^{\circ}49^{\prime}42^{\prime\prime}$&175&60&402&a,i\\
Sculptor\tablenotemark{b}&$01^{h}00^{m}09^{s}$&$-33^{\circ}42^{\prime}33^{\prime\prime}$&102&88&(0.234)&a,b,h,v,y\\
LGS3&$01^{h}03^{m}55^{s}$&$+21^{\circ}53^{\prime}06^{\prime\prime}$&-287&810&0.16&a,d,h,o,p\\
IC 1613&$01^{h}04^{m}54^{s}$&$+02^{\circ}08^{\prime}00^{\prime\prime}$&-237&700&54&a,q,r\\
Phoenix&$01^{h}51^{m}06^{s}$&$-44^{\circ}26^{\prime}41^{\prime\prime}$&-13&445&0.17&a,b,l\\
Fornax\tablenotemark{b}&$02^{h}39^{m}59^{s}$&$-34^{\circ}26^{\prime}57^{\prime\prime}$&53&138&$(0.15)$&a,h,z\\
LMC\tablenotemark{a}&$05^{h}23^{m}34^{s}$&$-69^{\circ}45^{\prime}24^{\prime\prime}$&324&50&500&a,b,i\\
Carina&$06^{h}41^{m}37^{s}$&$-50^{\circ}57^{\prime}58^{\prime\prime}$&224&101&$<0.005$&a,h,aa\\
Leo A&$09^{h}59^{m}24^{s}$&$+30^{\circ}44^{\prime}42^{\prime\prime}$&22.3&690&8&a,k\\
Sextans B&$10^{h}00^{m}00^{s}$&$+05^{\circ}19^{\prime}42^{\prime\prime}$&300&1345&45&a,m,r\\
Antlia&$10^{h}04^{m}04^{s}$&$-27^{\circ}19^{\prime}52^{\prime\prime}$&351&1235&0.72&a,b,e\\
Leo I&$10^{h}08^{m}28^{s}$&$+12^{\circ}18^{\prime}23^{\prime\prime}$&286&250&$<0.03$&a,h,aa\\
Sextans A&$10^{h}11^{m}06^{s}$&$-04^{\circ}42^{\prime}30^{\prime\prime}$&328&1440&78&a,s,w\\
Sextans&$10^{h}13^{m}03^{s}$&$-01^{\circ}36^{\prime}53^{\prime\prime}$&140&86&$<0.004$&a,h,aa\\
Leo II&$11^{h}13^{m}29^{s}$&$+22^{\circ}09^{\prime}17^{\prime\prime}$&76&205&$<0.02$&a,v,aa\\
GR8&$12^{h}58^{m}40^{s}$&$+14^{\circ}13^{\prime}00^{\prime\prime}$&214&1590&4.5&a,p,t,r\\
Ursa Minor&$15^{h}09^{m}08^{s}$&$+67^{\circ}13^{\prime}21^{\prime\prime}$&-247&66&$<0.011$&a,v,aa \\
Draco&$17^{h}20^{m}12^{s}$&$+57^{\circ}54^{\prime}55^{\prime\prime}$&-293&82&$<0.017$&a,aa\\
Sagittarius&$18^{h}55^{m}03^{s}$&$-30^{\circ}28^{\prime}42^{\prime\prime}$&140&24&$<0.0003$&a,u,aa\\
SagDIG&$19^{h}29^{m}59^{s}$&$-17^{\circ}40^{\prime}41^{\prime\prime}$&-75&1060&8.8&a,d,p,n\\
DDO 210&$20^{h}46^{m}52^{s}$&$-12^{\circ}50^{\prime}53^{\prime\prime}$&-141&800&1.9&a,b,p\\
IC 5152&$22^{h}02^{m}42^{s}$&$-51^{\circ}17^{\prime}42^{\prime\prime}$&122&1590&67&a,m,x\\
Tucana&$22^{h}41^{m}50^{s}$&$-64^{\circ}25^{\prime}10^{\prime\prime}$&182&880&$<0.39$&a,b,c,h,aa\\
UGCA 438&$23^{h}26^{m}27^{s}$&$-32^{\circ}23^{\prime}18^{\prime\prime}$&62&1320&6.2&a,m,n\\
PegDIG&$23^{h}28^{m}36^{s}$&$+14^{\circ}44^{\prime}35^{\prime\prime}$&-183&760&3.4&b,g\\
\enddata
\label{Table 3}
\tablecomments{References: a: \cite{mateo98} and references therein, b: \cite{grebel03} and references therein, c: \cite{tolstoy04}, d: \cite{young97}, e: \cite{tolstoy00}, f: \cite{lewis07}, g: \cite{huchra99}, h: \cite{BCS06}, i: \cite{bruns05}, j: \cite{humason56}, k: \cite{brown07}, l: \cite{irwin02}, m: \cite{huchtmeier86}, n: \cite{longmore82}, o: \cite{thuan79}, p: \cite{lo93}, q: \cite{lake89}, r: \cite{hoffman96}, s: \cite{huchtmeier88}, t: \cite{carignan90}, u: \cite{koribalski94}, v: \cite{knapp78}, w: \cite{skillman88}, x: \cite{BR00}, y: \cite{carignan98}, z: \cite{BCS06}, aa: This paper. }
\tablenotetext{a}{\scriptsize The LMC and SMC are included here for reference, but are not included in the figures.}
\tablenotetext{b}{\scriptsize The HI mass given is that of the nearby HI cloud which may or may not be associated with the galaxy.}
\end{deluxetable*}
\tabletypesize{\scriptsize}
\begin{deluxetable*}{llllll}
\tablewidth{0pt} 
\tablecaption{Orbital Characteristics and Hot Halo Densities}  
\tablehead{\colhead{Satellite}  &\colhead{Most Likely} &\colhead{Range in}& \colhead{$n_{peri}$} & \colhead{Range in $n_{peri}$}& \colhead{V$_{peri}$}\\
\colhead{}&\colhead{Perigalacticon (kpc)}&\colhead{Perigalacticons (kpc)}&\colhead{($\times$ 10$^{-4}$ cm$^{-3}$)}&\colhead{($\times$ 10$^{-4}$ cm$^{-3}$)}&\colhead{(km s$^{-1}$)}}
\startdata 
Carina&20&3-63&0.85&0.55 - 3.9&443\\
Ursa Minor&40&10-76&2.1&0.13 - 7.2&283\\
Sculptor&68&31-83&2.7&0.51 - 3.9&251\\
Fornax&118&66-144&3.1&0.98 - 4.6&231\\
\enddata
\label{Table 4}
\tablecomments{The range in perigalacticons are the 90\% confidence level for the orbits listed in the references in the text and the range in densities correspond to that range of perigalacticons. $n_{peri}$ and V$_{peri}$ are the density and velocity at the most likely perigalacticon.}
\end{deluxetable*}

\clearpage

\end{document}